\documentclass[12pt]{iopart}

\usepackage{inputenc}
\usepackage{amsfonts}
\usepackage{bbm}
\usepackage[printonlyused,smaller]{acronym}
\usepackage{graphicx}
\usepackage{appendix}

\usepackage[section]{algorithm}
\usepackage{algpseudocode}
\usepackage{hyperref}

\usepackage{xcolor}
\input{tudcolours.def}

\usepackage[capitalise,english]{cleveref}

\newcommand{\setOfReals}{\mathbb{R}}
\newcommand{\setOfNaturals}{\mathbb{N}}

\newcommand{\setOfPositiveReals}{{\setOfReals_{+}}}

\newcommand{\borel}[1]{\mathcal{B} (#1 )}

\newcommand{\sys}[1]{\textsc{#1}}

\newcommand{\measureIntegral}[2]{\langle#1, #2 \rangle }

\makeatletter
\let\oldabs\abs
\def\abs{\@ifstar{\oldabs}{\oldabs*}}

\newcommand{\cadlag}{c\`adl\`ag\,}

\newcommand{\reactionrate}[1]{\;{\longrightarrow}^{#1}\;}

\newcommand{\defeq}{:=}

\newcommand{\indicator}[1]{\mathsf{1}_{#1}}

\newcommand{\differential}[1]{\,\mathrm{d} #1}
\newcommand{\timeDerivative}[1]{\frac{\differential}{\differential t} #1 }
\newcommand{\partialDerivative}[2]{\frac{\partial}{\partial #1} #2 }

\newcommand{\eqstop}{.}
\newcommand{\eqcomma}{,}

\newcommand{\prob}{\mathsf{P}}
\newcommand{\probOf}[1]{\prob\left(#1\right)}

\newcommand{\ie}{\textit{i.e.}}
\newcommand{\eg}{\textit{e.g.}}

\renewcommand {\theequation}{\arabic{section}.\arabic{equation}}

\begin{document}
\title[Incorporating age and delay into models for biophysical systems]{Incorporating age and delay into models for biophysical systems}

\author{Wasiur R. KhudaBukhsh\hspace{1.5mm}\href{https://orcid.org/0000-0003-1803-0470}{\includegraphics[width=3mm]{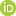}}$^1$, Hye-Won Kang\hspace{1.5mm}\href{https://orcid.org/0000-0002-7350-5355}{\includegraphics[width=3mm]{ORCID-iD_icon-16x16}}$^2$, Eben Kenah\hspace{1.5mm}\href{https://orcid.org/0000-0002-7117-7773}{\includegraphics[width=3mm]{ORCID-iD_icon-16x16}}$^3$, and Grzegorz A. Rempa{\l}a\hspace{1.5mm}\href{https://orcid.org/0000-0002-6307-4555}{\includegraphics[width=3mm]{ORCID-iD_icon-16x16}}$^4$}

\address{$^1$ Mathematical Biosciences Institute and the College of Public Health, The Ohio State University, 1735 Neil Avenue, Columbus OH 43210, USA}

\address{$^2$ Department of Mathematics and Statistics, University of Maryland, Baltimore County, 1000 Hilltop Circle, Baltimore MD 21250, USA}

\address{$^3$ Division of Biostatistics, College of Public Health, The Ohio State University, 1841 Neil Avenue, Columbus OH 43210, USA}

\address{$^4$ Mathematical Biosciences Institute and the College of Public Health, The Ohio State University, 1735 Neil Avenue, Columbus OH 43210, USA}

\ead{khudabukhsh.2@osu.edu}




\begin{abstract}
In many biological systems, chemical reactions or changes in a physical state  are assumed to occur instantaneously. For describing the dynamics of those systems, 
Markov models that require exponentially distributed inter-event times have been used widely. However, some biophysical
processes such as gene transcription and translation are known to have a significant gap between the initiation and the completion of the processes, which  renders the usual assumption 
of exponential distribution untenable. 
In this paper, we consider relaxing this assumption by incorporating age-dependent random time delays into the system dynamics. 
We do so by constructing a measure-valued Markov process on a more abstract state space, which allows us to keep track of 
the ``ages'' of molecules participating in a chemical reaction. 

We study the large-volume limit of such age-structured systems. We show that, when appropriately scaled, the
stochastic system can be approximated by a system of \acp{PDE} in the large-volume limit, as opposed to \acp{ODE} in the classical theory. We show how the 
limiting \ac{PDE} system can be used for the purpose of further model reductions and for devising efficient simulation algorithms. In order to 
describe the ideas, we use a simple transcription process as a running example. We, however, note that the methods developed in this paper 
apply to a wide class of biophysical systems. 
\end{abstract}


  \maketitle



\section{Introduction}\label{sec:intro}

We consider biophysical systems described by a set of chemical reactions. The chemically identical molecular entities in the system are called (chemical) 
species. A chemical reaction refers to the event of creation, annihilation, or conversion of a number of molecules of one or more species. Here, we 
assume the system is well mixed spatially in that a randomly chosen molecule of a species has an equal chance to chemically interact 
with any other molecule of any species in the system. A \ac{CTMC} is a natural choice to model the species copy numbers of  
such systems. 

When modelling \acp{CRN} stochastically using \acp{CTMC}, one assumes that every reaction occurs instantaneously after an 
exponentially distributed amount of time. Whenever a reaction takes place, we update the system state. A random time-change 
representation of the Poisson process is often used to write the trajectory equations and to analyze the system dynamics~\cite{Ball:2006:AAM,Anderson:2011:CTM,Kang:2013:STM,Kang:2012:MAH}. The sample paths
of the \ac{CTMC} are  simulated exactly using the Doob--Gillespie's \ac{SSA}~\cite{Gillespie:1976:GMN,Gillespie:1977:ESS,Gillespie:2007:SSC} or the next reaction method by Gibson and Bruck~\cite{Gibson:2000:EES}.




\subsection{Delays are inherent and a useful model reduction tool}
It has been reported that some biological processes do not take place instantaneously. In other words, there 
is a time lag between the initiation and the completion of the process. Time delays are observed 
inherently in many biological systems, such as gene transcription~\cite{Palangat:1998:TPH,Hoyle:2013:TPE,Swinburne:2008:IDT} and translation~\cite{Baron:2019:IND}, cell 
cycle in cancer treatment~\cite{Zhou:2002:TRM}, intracellular viral dynamics~\cite{Herz:1996:VDL,Bai:2019:EDV}, control of 
infectious diseases~\cite{Fraser:2004:FMI}, population growth~\cite{Kuang:1993:DDE,Giang:2005:DEM}, RNA and protein folding~\cite{vandenBerg:2000:MCP,vanMeerten:2001:TCD}, and enzyme catalyzed reactions~\cite{Easterby:1981:GTT,Kekenes:2015:ELC}.
Sometimes time delays are  introduced purposefully as a useful means to reduce model complexity and compensate for the lack of 
experimental observation in both deterministic and stochastic models of biological processes.

Unimportant processes or unobserved reactions can be replaced by  time delays. For example, production of hes1 mRNA from hes1 gene has been modeled using delay differential 
equations where detailed mRNA synthesis and processing steps are replaced by a time delayed reaction~\cite{Monk:2003:OEH}. While
modeling the mammalian circadian clock, intermediate protein dynamics can be simplified as transcriptional feedback loops with time delayed 
variables in delay differential equations~\cite{Korencic:2012:ICE}. In enzyme catalyzed reactions with 
multiple intermediates, the production of the final product can be expressed as a distributed delay equation, which is a useful tool when  measurements on multiple 
intermediates in the experiment are not available~\cite{Hinch:2004:MEE}.

Introduction of  time delays as a model reduction technique has also been applied in  discrete stochastic models 
for \acp{CRN}. For instance, model complexity of unimolecular reaction networks is reduced 
by generating delay distributions with key model features that are derived by computing first passage times of 
target species~\cite{Barrio:2013:RCR}. In ~\cite{Choi:2020:BID}, the production of yellow fluorescent protein has been described using a time-delayed birth and death process where a randomly distributed time delay was generated to simplify a sequence of steps in gene activation.

\subsection{Our contribution}
In a majority of previous works, the focus was on investigating stochastic models for \acp{CRN} with constant 
or randomly distributed time delays. In those models, the probability that a reaction occurs within the next short 
interval of time is commonly described by a propensity (also known as intensity) function of the reaction. The waiting time 
for non-delayed reactions is exponentially distributed~\cite{Hepp:2015:AHS}.
In practice, the occurrence of some reactions is not only determined by the molecular counts of the 
reactants but also affected by the age distributions or lifetimes of the reactant molecules. For example, mRNA 
decay rates vary depending on the age of each mRNA. Moreover, the age of the mRNAs determines polysome size distributions and protein 
synthesis rates in translation (\cite{Valleriani:2010:TMR,Deneke:2013:CDP}, Chapters 3 and 5 in \cite{Deneke:2012:TMD}). It was also 
reported that an mRNA tail length distribution depends on the average age of mRNA population and that the tail-length distribution plays a significant role in deadenylation and decay dynamics of mRNA populations~\cite{Prieto:2000:GTM,Eisen:2020:DCM}. 

When time delays are used to aggregate out unimportant processes and reduce model complexity, it makes more sense that 
the length of 
time delay depends on the age of each reactant molecule (\eg,  mRNA, protein, and enzymes). Therefore, it is worthwhile to consider an 
individual-based age-structured stochastic model for \acp{CRN}.

In this work, we develop a way to describe \acp{CRN} with random time delays and non-delayed reaction rates 
incorporating the {\textit{age}} of each reactant and making use of {\textit{hazard functions}} in 
survival analysis~\cite{KhudaBukhsh:2019:SDS,Calderazzo:2019:FIS}. In our approach, the hazard functions
 are set as constant, time-dependent, or age-dependent functions generalizing the notion of reaction rate constants 
 in propensity functions. Our model keeps track of the age of each reactant molecule and provides a new 
 way to express time delays in non-Markovian models. Moreover, the method also allows
  us to describe discrete stochastic \acp{CRN} with constant or randomly-distributed time delays without 
  age dependence, as considered in  previous works. We study the large-volume limit of the proposed non-Markov \ac{CRN} and 
  provide a mean-field \ac{PDE} limit for the age densities  by virtue of the \ac{LLN}, as opposed to an \ac{ODE} limit 
  in the classical theory. We show how the \ac{PDE} limit can be used to approximate \acp{MFPT} in the context of 
  \acp{CRN}. As another by-product of the \ac{LLN}, we show how an efficient (fast) hybrid simulation algorithm 
  can be devised when a subset of the \ac{CRN} is abundantly available, giving a flavor of multiscale approximation. Finally, as simple applications of our approach,  we 
  briefly discuss a prokaryotic auto-regulation and the \ac{QSSA} in the context of  the Michaelis--Menten enzyme kinetic reactions. Numerical 
  examples have been provided wherever deemed necessary. For the sake of ready usage of our methods, the Julia scripts used in the numerical examples 
  have been made available via a GitHub repository \cite{delaymodel2020}.




The following notational conventions are adhered to throughout the paper. We 
use $\indicator{A}(x)$ to denote the indicator (or characteristic) function of 
a set $A$, \ie, $\indicator{A}(x) =1$ if and only if $x\in A$. Given a suitable  
space $E$, let $D([0,\infty), E)$ (or $ D([0,T], E)$) denote the space 
of $E$-valued \cadlag functions defined on $[0,\infty)$ (or $[0,T]$, for 
some $T >0$). The set of Borel subsets of a set $A$ will be denoted by $\borel{A}$.  The set of natural numbers 
are denoted by $\setOfNaturals$. 
The set of 
real numbers is denoted by $\setOfReals$. Other notations will be introduced as and when needed.


\section{The simplest model with a delay}
\label{section:simple_model}

Let us consider a simple \ac{CRN} with two chemical species $A$ and $B$. First, we shall describe the 
standard Markovian approach and then introduce an age structure to allow  non-exponential holding times. The following network describes 
the production and the degradation of $A$ along with a conversion from $A$ to $B$
\begin{equation}
  \eqalign{
    \emptyset \reactionrate{b}   A & \reactionrate{\tau} B \eqcomma \\
    A \reactionrate{d} \emptyset & \eqstop 
  }
  \label{eq:model0}
\end{equation}
where $b$, $\tau$, and $d$, depending on whether we are in the Markovian or non-Markovian setup, will be either reaction rate constants or hazard functions for 
the production of $A$, the conversion from $A$ to $B$, and the degradation of $A$, respectively. 

An example similar to the \ac{CRN} in  \Cref{eq:model0}  was  
investigated in some previous works with time delays~\cite{Bratsun:2005:DSO,Koyama:2013:ASM}. It is worth noting that the simplistic \ac{CRN} described in \Cref{eq:model0} can be thought of as a model reduction 
of a more complex \ac{CRN}. For instance, a series of conversion type reactions
\begin{equation*}
  A \reactionrate{k_1} A_1 \reactionrate{k_2} A_2 \reactionrate{k_3} \cdots \reactionrate{k_n} B    
\end{equation*}
can be described by a single 
conversion reaction $A \reactionrate{\tau} B$ with an appropriate associated hazard function $\tau$. Similarly, a 
series of birth-death-conversion type reactions
\begin{equation*}
  \eqalign{
    \emptyset \reactionrate{b} A \reactionrate{k_1}   A_1 \reactionrate{k_2} A_2 \reactionrate{k_3} \cdots \reactionrate{k_n} B \eqcomma  & \\
    A \reactionrate{d}   \emptyset,  A_1 \reactionrate{d_1}   \emptyset, A_2 \reactionrate{d_2}   \emptyset, \cdots, A_n \reactionrate{d_n}   \emptyset
  }
\end{equation*}
can be approximated by a single birth type reaction $ \emptyset \reactionrate{\tau}   B $ with an appropriate hazard 
function $\tau$. Therefore, even a simplistic model such as the \ac{CRN} in 
\Cref{eq:model0} covers a nontrivial class of \acp{CRN} and builds the foundation for studying more complex 
\acp{CRN}. 

\subsection{Standard Markov approach}\label{sec:model0_markov}
The standard way to 
model the  \ac{CRN} in  \Cref{eq:model0} is to use a \ac{CTMC} to describe the counts of molecules of the species $A$ 
and $B$ over time. In such a model, whenever each reaction fires, the consumption and the production of molecules 
are instantaneous. Let $\tilde{X}_A, \tilde{X}_{B}$ denote   the stochastic processes counting the 
copy numbers of the species $A$ and $B$ respectively. Here, the quantities $b, \tau$, and $d$ are reaction rate constants. The propensity functions corresponding to the three chemical reactions 
are defined as
\begin{eqnarray*}
\lambda_b(t)= b, \qquad \lambda_\tau(t)= \tau x_A(t), \qquad \lambda_d(t)= d x_A(t),
\end{eqnarray*}
where $x_i(t)$ denotes the number of molecules of the chemical species $i$ at time $t$, for $i=A,B$. Define $T_k$ to be the 
waiting time until the next reaction of type birth ($k=b$), conversion ($k=\tau$),
and death ($k=d$). Then, $T_k$ is exponentially distributed with rate $\lambda_k(t)$ for $k=b,\tau,d$. The probability 
of each reaction's occurrence is expressed in terms of the corresponding propensity function as follow:
\begin{eqnarray*}
\probOf{t\le T_k < t+\Delta t  \mid \tilde{X}_A(t) = x_A, \tilde{X}_B(t) = x_B  } &\approx& \lambda_k(t)\Delta t + o(\Delta t) 
\end{eqnarray*}
for $k=b,\tau,d$ 
when $\Delta t$ is small enough. Then, the trajectory equations can be written in a straightforward 
fashion following the random time changed representation of Poisson processes as 
\begin{equation*}
  \eqalign{
    \tilde{X}_A(t) & = \tilde{X}_A(0) + R_{1}\left(bt\right) - R_{2}\left(\int_0^t \tau \tilde{X}_A(s) \differential{s} \right) - R_{3}\left(\int_0^t d \tilde{X}_A(s) \differential{s} \right) \eqcomma \\
    \tilde{X}_{B}(t) & =  R_{2}\left(\int_0^t \tau \tilde{X}_A(s) \differential{s} \right) \eqcomma 
  }
\end{equation*}
where $R_1, R_2$, and $R_3$ are unit rate Poisson processes \cite{Anderson:2011:CTM}. We assume we do not have 
any $B$ molecules in the system initially, \ie, $\tilde{X}_b(0) = 0$.  Now, if we scale the 
stochastic processes by a scaling parameter $n$, e.g. volume of the system,  it follows directly from the 
\ac{LLN} for Poisson processes \cite{kurtz1970solutions,kurtz1978strong} that the scaled process $(n^{-1}\tilde{X}_A, n^{-1}\tilde{X}_B )$ can be 
approximated by the solution to the following system of \acp{ODE}:
\begin{equation*}
  \eqalign{
    \timeDerivative{x_A(t)} &= - (\tau + d) x_{A}(t) \eqcomma \\ 
    \timeDerivative{x_{B}(t)} &= \tau x_{A}(t) \eqstop  
  }
\end{equation*}
Notice that the birth rate $b$ vanishes in the limit because we did not assume any scaling of $b$ with respect to $n$. In general, one 
would assume that the overall birth rate scales linearly with $n$ so that it is sustained in the limit. 


\subsection{Age-structured model}
Now, let us introduce \emph{age} and \emph{delay} into the \ac{CRN} described by \Cref{eq:model0}. We assume that the production rate of $B$ and 
the  degradation rate of $A$ depend on the age of the reactant molecule of species $A$. We may define age of a molecule in many different ways. 
The most straightforward of them is the biological or the physical age, which we take as the time duration since the molecule was born 
or created. In systems where a certain reaction can fire only when a gene is activated, one could define age as the 
time duration since activation of the gene. In some cases, it may be desirable to define delays in terms of time duration since the 
initiation of a reaction. The notion of age is sufficiently general to 
account for those cases as well. For example, a reaction $A \rightarrow B$ in which the delay is defined 
purely in terms of time difference between initiation and completion of the reaction, can be replaced by the reaction system 
$A \rightarrow F \rightarrow B$ where $F$ is a fictitious species. The physical age of this fictitious species $F$ 
is precisely the time since the initiation of the reaction $A \rightarrow B$. Now, putting an appropriate hazard function on the 
reaction $F \rightarrow B$, we can introduce a random or a deterministic delay in the reaction $A\rightarrow B$. Therefore, for the \ac{CRN} in \Cref{eq:model0}, it seems 
sufficient to define the age to be the physical age of the 
molecules of $A$. 

When we have an age-structured model, the counts (copy numbers of the species $A$, and $B$) are 
inherently non-Markovian unless the holding times are exponentially 
distributed. However, if we keep track of the ages of the molecules 
in addition to the counts, we can get a Markov system, albeit on a 
more abstract state space. A neat way to do so is to use 
measure-valued processes that keep track of the age distribution 
of the molecules over time. Moreover, the measure-valued processes 
are also Markovian, which allows us to make use of the  
already existing limit theory for Banach space-valued Markov processes. This 
approach to age-structured modeling in biology is not new. Our work 
builds on the existing literature \cite{Fournier2004microscopic,Tran2008limit,Tran2009traits,Meleard:2012:SFS}. In the 
next section, we describe how the measure-valued processes 
can be utilized in the context of the \ac{CRN} in \Cref{eq:model0}.

\subsection{The measure-valued process and the limiting system}
\label{section:simple_model_limit}


Let us denote by $N_A(t)$ and $N_B(t)$ the numbers of molecules 
of the chemical species $A$ and $B$ at time $t$. Then, individual 
molecules of $A$ are labelled $1,2,\cdots,N_A(t)$ following some partial order (\eg, increasing order of age). We denote 
the age of the $i$-th molecule of the species $A$ by $a_i(t)$ for $i=1,2,\cdots,N_A(t)$. 
Similarly, we denote by $b_j(t)$ the age of the $j$-th molecule 
of the species $B$ at time $t$. 
Now, we define a measure-valued process $X_t=\left(X_t^A,X_t^B\right)$ where $X_t^A$ and $X_t^B$ describe the 
age distributions of chemical species $A$ and $B$ at time $t$. To be more precise,  we define
\begin{equation}
  X_t^A \defeq \sum_{i=1}^{N_A(t)} \delta_{a_i(t)} \eqcomma 
  \quad X_t^B \defeq  \sum_{i=1}^{N_B(t)} \delta_{b_i(t)} \eqcomma  
\end{equation}
where $\delta_x$ is the Dirac delta function, which takes value $1$  if the argument to the function is $x$ and 
zero otherwise. 
The components $X_t^A$, and $X_t^B$ of $X_t$ 
are finite point measures with atoms placed on the individual 
ages of the molecules. For example, $X_t^A \left( (0.5, 11.25] \right) = \sum_{i=1}^{N_A(t)} \delta_{a_i(t)} \left( (0.5, 11.25] \right)$ gives us 
the count of species $A$ molecules with ages in the set $(0.5, 11.25]$ at time $t$. In 
general, $X_t^A \left(F\right)$ gives us the count of species $A$ molecules whose ages lie in the set $F$ at time $t$.

For any point measure $\mu=\sum_{i=1}^n\delta_{x_i}$ and a measurable function $f$, we express the 
integration of the function $f$ with respect to the measure $\mu$ as
\begin{eqnarray*}
\langle\mu,f\rangle &\defeq& \int f\,d\mu = \sum_{i=1}^n f(x_i).
\end{eqnarray*}
Therefore, we have $N_{A}(t) = \langle X_t^A,1\rangle = X_t^A \left( \setOfPositiveReals \right)$ and 
$N_{B}(t) = \langle X_t^B,1\rangle = X_t^B \left( \setOfPositiveReals \right)$ where $1$ stands for the identity function. The process 
$X$ is a Markov process on the space $D([0,T], \mathcal{M}_{P}(\setOfPositiveReals) \times \mathcal{M}_{P}(\setOfPositiveReals) )$
where $T>0$ is a finite time horizon and $\mathcal{M}_{P}(\setOfPositiveReals)$ is the space of finite, point measures 
on $\setOfPositiveReals$, the set of non-negative real numbers. 

In order to simplify notations, we introduce maps $\sigma_i:  \mathcal{M}_{P}(\setOfPositiveReals) \rightarrow \setOfPositiveReals$, for 
$i = 1, 2, 3, \ldots$, 
the purpose of which is to extract the $i$-th atom  (the age of the $i$-th molecule) from a point measure following 
some partial order (\eg, ``greater or equal to'' relation). Therefore, $\sigma_{i}(X_t^A)$ gives us the age of the 
$i$-th molecule of the species $A$ at time $t$. We can now write down the trajectory equations:
\begin{eqnarray*}
  \eqalign{
    X_t^A & = \sum_{k=1}^{N_A(0)} \delta_{t+ \sigma_k(X_0^A)} + \int_0^t \int_0^\infty \delta_{t-s}\, \indicator{\theta \leq b}\,  Q_1(ds, d\theta) \\
    & \quad  - 
    \int_0^t \int_{\setOfNaturals} \int_0^\infty  \delta_{t-s+  \sigma_i(X_{s-}^A)} \, \indicator{i \leq N_A(s-)}\, \indicator{\theta \leq \tau (\sigma_i (X_{s-}^A)) }\, Q_2(ds, di, d\theta) \\
    & \quad  - 
    \int_0^t \int_{\setOfNaturals} \int_0^\infty  \delta_{t-s+  \sigma_i(X_{s-}^A)} \, \indicator{i \leq N_A(s-)}\, \indicator{\theta \leq d (\sigma_i (X_{s-}^A)) }\, Q_3(ds, di, d\theta) \eqcomma 
  }  \\
  \eqalign{
    X_t^B & =  
    \int_0^t \int_{\setOfNaturals} \int_0^\infty  \delta_{t-s} \, \indicator{i \leq N_A(s-)}\, \indicator{\theta \leq \tau (\sigma_i (X_{s-}^A)) }\, Q_2(ds, di, d\theta)  \eqcomma 
    }
\end{eqnarray*}
where $Q_1, Q_2$, $Q_3$ are independent \acp{PPM} with intensity measures $\differential{s}\times \differential{\theta}$, 
$\differential{s}\times \differential{i}\times \differential{\theta}$, and $\differential{s}\times \differential{i}\times \differential{\theta}$ respectively, where 
$\differential{i}$ is a counting measure on $\setOfNaturals$, and $\differential{s}$ and $\differential{\theta}$ 
are Lebesgue measures on $\setOfPositiveReals$. In order to study moments and martingale properties of $X_t^A$ and 
$X_t^B$, it is worthwhile to check that
\begin{eqnarray*}
  \eqalign{
    \measureIntegral{X_t^A}{f_t} & = \sum_{k=1}^{N_A(0)} f_t({t+ \sigma_k(X_0^A)}) + \int_0^t \int_0^\infty f_t({t-s})\, \indicator{\theta \leq b}\,  Q_1(ds, d\theta) \\
    & \quad  - 
    \int_0^t \int_{\setOfNaturals} \int_0^\infty  f_t({t-s+  \sigma_i(X_{s-}^A)}) \, \indicator{i \leq N_A(s-)}\, \indicator{\theta \leq \tau (\sigma_i (X_{s-}^A)) }\, Q_2(ds, di, d\theta) \\
    & \quad  - 
    \int_0^t \int_{\setOfNaturals} \int_0^\infty  f_t({t-s+  \sigma_i(X_{s-}^A)}) \, \indicator{i \leq N_A(s-)}\, \indicator{\theta \leq d (\sigma_i (X_{s-}^A)) }\, Q_3(ds, di, d\theta) \eqcomma 
  }  \\
  \eqalign{
    \measureIntegral{X_t^B}{f_t} & =  
    \int_0^t \int_{\setOfNaturals} \int_0^\infty  f_t({t-s}) \, \indicator{i \leq N_A(s-)}\, \indicator{\theta \leq \tau (\sigma_i (X_{s-}^A)) }\, Q_2(ds, di, d\theta)  \eqcomma 
    }
\end{eqnarray*}
for a sufficiently large class of test functions $f: (a, s) \rightarrow f_s(a) $.

As in the standard Markov model in \Cref{sec:model0_markov}, we are now interested in the large-volume limit ($n \rightarrow \infty$) of the scaled stochastic process $n^{-1} X$. Under reasonable assumptions on the 
hazard functions and the initial age distribution of the $A$ molecules, one would expect the scaled process $n^{-1} X_t$
to converge to a deterministic function $x_t \defeq (x_t^A, x_t^B)$ whose components $x^A$, and $x^B$ are themselves measure-valued 
functions. This is by virtue of the \ac{LLN}. Since we have 
\begin{equation*}
  f_t(a+t-s) = f_s(a) + \int_s^t \left( \partialDerivative{u}{ f_{u}(a+u-s)} + \partialDerivative{a}{f_{u}(a+u-s) } \right) \differential{u}  \eqcomma
\end{equation*}
we would  expect the limit $x$ to satisfy
\begin{eqnarray*}
  \eqalign{
    \measureIntegral{x_t^A}{f_t} & = \measureIntegral{x_0^A}{f_0} + \int_0^t \int_0^\infty \left( \partialDerivative{a}{f_s(a)} + \partialDerivative{s}{f_s(a)} - f_s(a) (\tau(a) + d(a))  \right) x_s^A(\differential{a}) \differential{s} \\
  }  \\
  \eqalign{
    \measureIntegral{x_t^B}{f_t} & =  \int_0^t \int_0^\infty \left( \partialDerivative{a}{f_s(a)} + \partialDerivative{s}{f_s(a)} + f_s(0) \tau(a)  \right) x_s^A(\differential{a}) \differential{s}  \eqstop
    }
\end{eqnarray*}
The convergence of the scaled stochastic process $n^{-1}X$ to the deterministic function $x$ can be proved  using techniques similar to those in 
\cite{Fournier2004microscopic,Champagnat2008individual,Tran2008limit,Tran2009traits,Meleard:2012:SFS}. A rigorous proof of the 
convergence and analytic properties of the limit in the context of a general non-Markovian \ac{CRN} will be provided elsewhere. 

Since the measure-valued function $x_t^B$ is determined entirely by $x_t^A$, it suffices to study $x_t^A$. The 
densities $y_A(t,a)$ of the measure $x_t^A$, when they exist, are an important quantity describing the distribution 
of age of the species $A$ molecules in the large-volume mean-field limit. The density function $y_A$ should satisfy 
\begin{equation}
  \left(\partial_{t} + \partial_{s}\right) y_{A}(t,s) = - \left(\tau(s) + d(s) \right)  y_{A}(t,s) \eqcomma 
  \label{eq:model0_pde}
\end{equation} 
with the initial and the boundary conditions 
\begin{eqnarray*}
  y_{A}(0,s) = f_A(s) \eqcomma \quad
  y_{A}(t,0) = 0 
  \eqcomma
\end{eqnarray*}
where $f_A(s) $ specifies the  age distribution of $A$ molecules at time $t=0$. To be more precise, it is the density 
 of the limiting measure $x_0^A$, which we assume exists, with respect to the Lebesgue measure.  Notice that the 
birth rate $b$ vanishes in the limit, as before, because we did not assume any scaling of the birth rate with respect to $n$. 

Let $y_B$ denote the limiting proportion of $B$ molecules in the system. Then, $y_B$ can be described entirely 
in terms of the density $y_A$ as a solution to the \ac{ODE}:
\begin{equation}
  \timeDerivative{y_{B}(t)} = \int_0^\infty \tau(s) y_{A}(t,s) \differential{s} \eqcomma
  \label{eq:model0b_ode}
\end{equation}
with the initial condition $y_B(0) = 0$. 
%
Luckily, the limiting system \Cref{eq:model0_pde} can be solved explicitly using standard analysis techniques:
\begin{equation*}
  y_{A}(t,s)  = f_{A}(s-t) S_{\tau}(s) S_{d}(s)/\left( S_{\tau}(s-t) S_{d}(s-t)\right) \eqcomma 
\end{equation*}
where $S_{\tau}$, and $S_{d}$ are the survival functions of the probability distributions characterized by the hazard functions $\tau$ and $d$ respectively. Therefore, the limiting concentration of $B$ molecules can be described by
\begin{equation*}
  y_{B}(t) = {\int_0^t \int_0^\infty \tau(v) y_{A}(u,v) \differential{v} \differential{u}} \eqstop
\end{equation*}

\begin{figure}[t]
\centering
\includegraphics[width=0.99\linewidth]{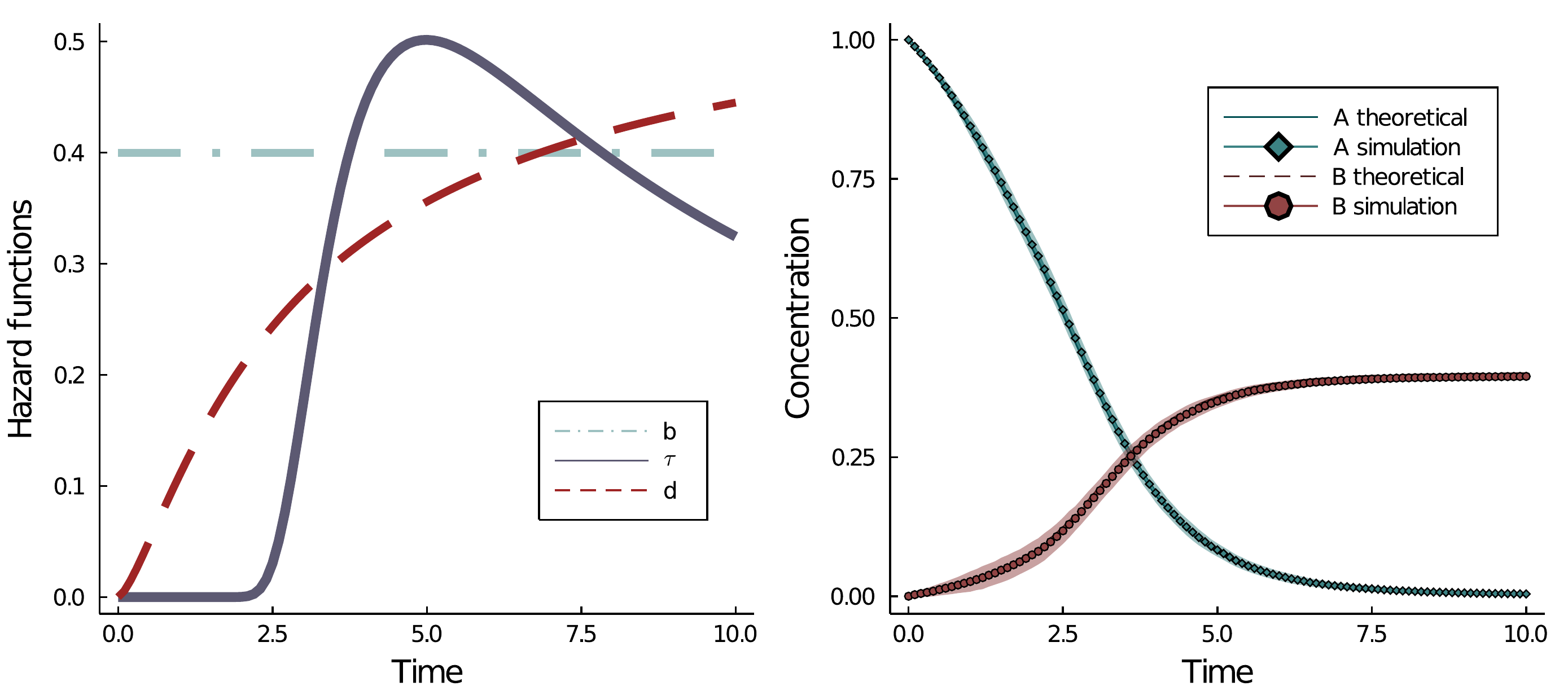}
  \caption{\label{fig:lln}%
  (Left) The shapes of the three hazard functions in the \ac{CRN} described by \Cref{eq:model0}. Here, $b=0.4$. The hazard functions $\tau$ and $d$ 
  correspond to a Generalized Extreme Value distribution with parameters $(1.25/0.30,1.250,0.30)$ and a Gamma distribution with 
  parameters $(2.5, 1.75)$ respectively. Here, the conversion reaction has been explicitly made a delayed one. (Right) Comparison 
  of the theoretical limiting trajectory and the simulated trajectories. Here, $n=500$, \ie, the initial 
  number of $A$ molecules is $500$.  It is evident that the theoretical limit 
  provides a fairly accurate approximation to the scaled processes.}
\end{figure}

In \Cref{fig:lln}, we numerically show the agreement between the theoretical limits in 
\Cref{eq:model0_pde,eq:model0b_ode}, and the stochastic simulation. As it can be verified, the approximation 
error vanishes in the limit. Because $X_t$ is a Markov process, the simulation of the stochastic \ac{CRN} in \Cref{eq:model0} can be carried out by adapting
the Doob--Gillespie's \ac{SSA}, which involves simulating two quantities at each step: 1) simulating the next 
reaction time; and 2) determining the reaction type. Note that, for the \ac{CRN} in \Cref{eq:model0}, there are 
$(2N_A(t) +1 )$ different reactions possible at time $t$, even though there are only three types of reactions. The 
next reaction time can be simulated by drawing an exponential random variable with rate equal to the 
total hazard (the
sum of the hazards corresponding to those $(2N_A(t) +1 )$ possible 
reactions). The type of reaction is then decided by drawing a categorical random variable whose probability masses 
are the ratios of the individual hazards and the total hazard. This discrete event simulation algorithm is a straightforward
adaptation of   Doob--Gillespie's \ac{SSA} for \acp{CTMC}. However, it must be noted that the simulation of a non-Markovian \ac{CRN} 
is computationally more expensive than the \acp{CTMC}.

In \Cref{sec:intro}, we mentioned that introduction of delay into a \ac{CRN} could also serve the purpose of 
model reduction. Indeed, the \ac{LLN} limit $y \defeq (y_A, y_B)$ provides a model reduction of the original 
non-Markovian \ac{CRN}. In the following, we shall discuss two other examples of usefulness of the \ac{LLN} limit in the 
form of a \ac{PDE} system. The first one approximates  \acp{MFPT}, while the second one describes a faster 
simulation algorithm.


\begin{figure}[t]
  \centering
  \includegraphics[width=0.99\linewidth]{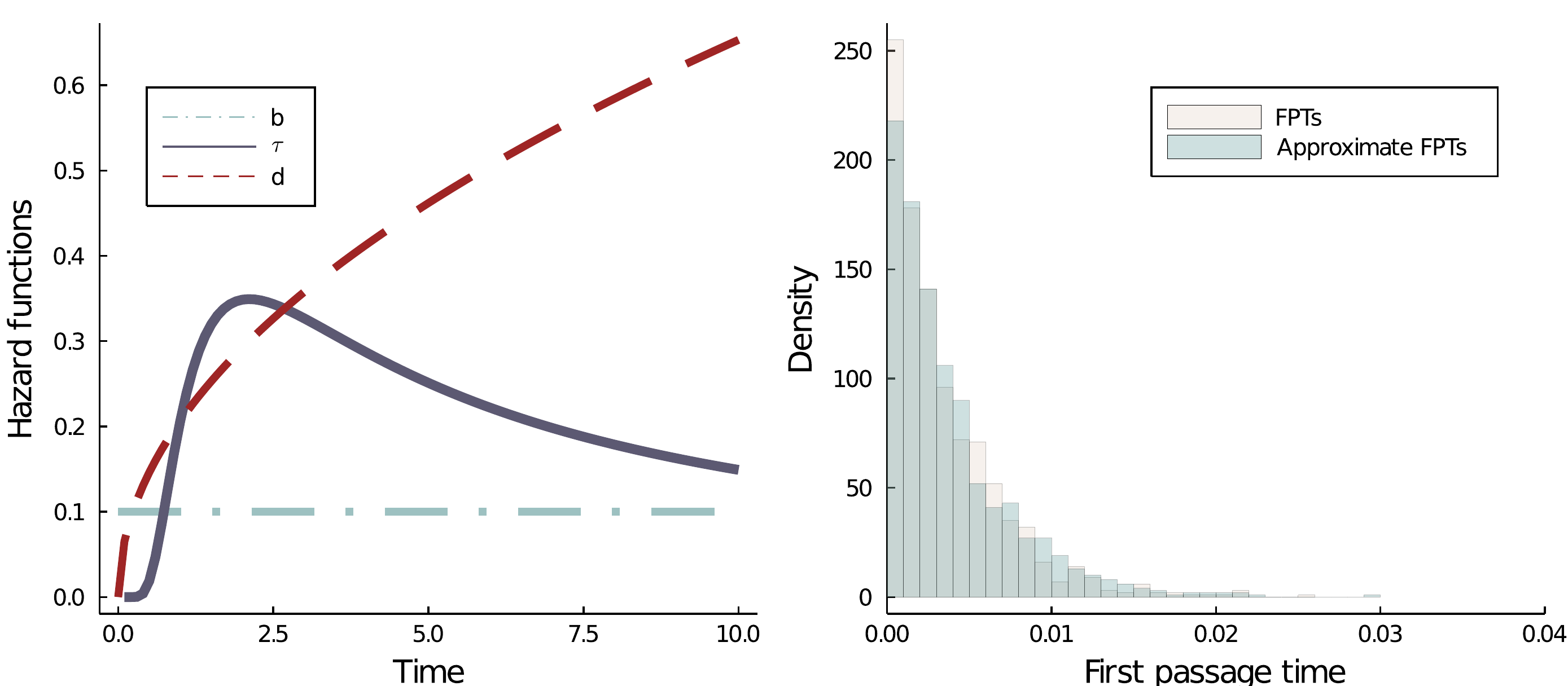}
  \caption{\label{fig:mfpt}%
  (Left) The shapes of the three hazard functions in the \ac{CRN} described by \Cref{eq:model0}. Here, $b=0.1$. The hazard functions $\tau$ and $d$ 
  charactertize an Inverse Gamma distribution with parameters $(1.75, 4.25)$ and a Weibull distribution 
  with parameters $(1.5, 3.75)$ respectively. (Right) The density of approximate \acp{FPT} match that of the true \acp{FPT}. Here, $n=2000$.}
  \end{figure}

\subsection{\aclp{MFPT}}
\aclp{MFPT} are  important quantities in the study of stochastic processes and dynamical systems. In the context 
of \acp{CRN}, they could arise in several ways \cite{MacNamara2008Multiscale,Jinsu2020slack}. For instance,  natural questions that could arise for the \ac{CRN} in 
\Cref{eq:model0_pde} are how long it takes to deplete all molecules of species $A$  or 
to produce the first molecule of $B$. One of the benefits of the \ac{LLN} limit is that it can be used to 
approximate \acp{FPT} when the scaling parameter~$n$ is sufficiently large. The following illustrates this point. 

Suppose we are interested in the time required to produce the 
first molecule of $B$. Then, for a sufficiently large $n$, the \ac{MFPT} can be approximated by 
\begin{equation}
  m = \left(\int_0^\infty \tau(s)\, n* y_A(0,s) \differential{s} \right)^{-1} \eqcomma  
  \label{eq:mfpt}
\end{equation}
which, of course, vanishes in the limit of $n\rightarrow \infty$. Moreover, the \acp{FPT} can be approximated by a random variable following an 
exponential distribution with mean $m$, whenever $n$ is sufficiently large. It follows that we can use a simple likelihood function (based on the 
exponential distribution) for the purpose of statistical 
inference of the underlying parameters, provided we have observations on the \acp{FPT}. This method, called dynamic survival analysis, of estimating 
parameters based on timings rather than counts  was recently explored in the context of an epidemiology in 
\cite{KhudaBukhsh:2019:SDS}. Dynamic survival analysis of general \acp{CRN} will be discussed in a future publication.

In \Cref{fig:mfpt}, we show 
the accuracy of this approximation when $n=2000$. The approximation appears to be reasonably accurate. More importantly, this suggests 
we might be able to devise an efficient simulation algorithm using such approximate results. We explore this idea next.


\begin{figure}[t]
  \centering
  \includegraphics[width=0.99\linewidth]{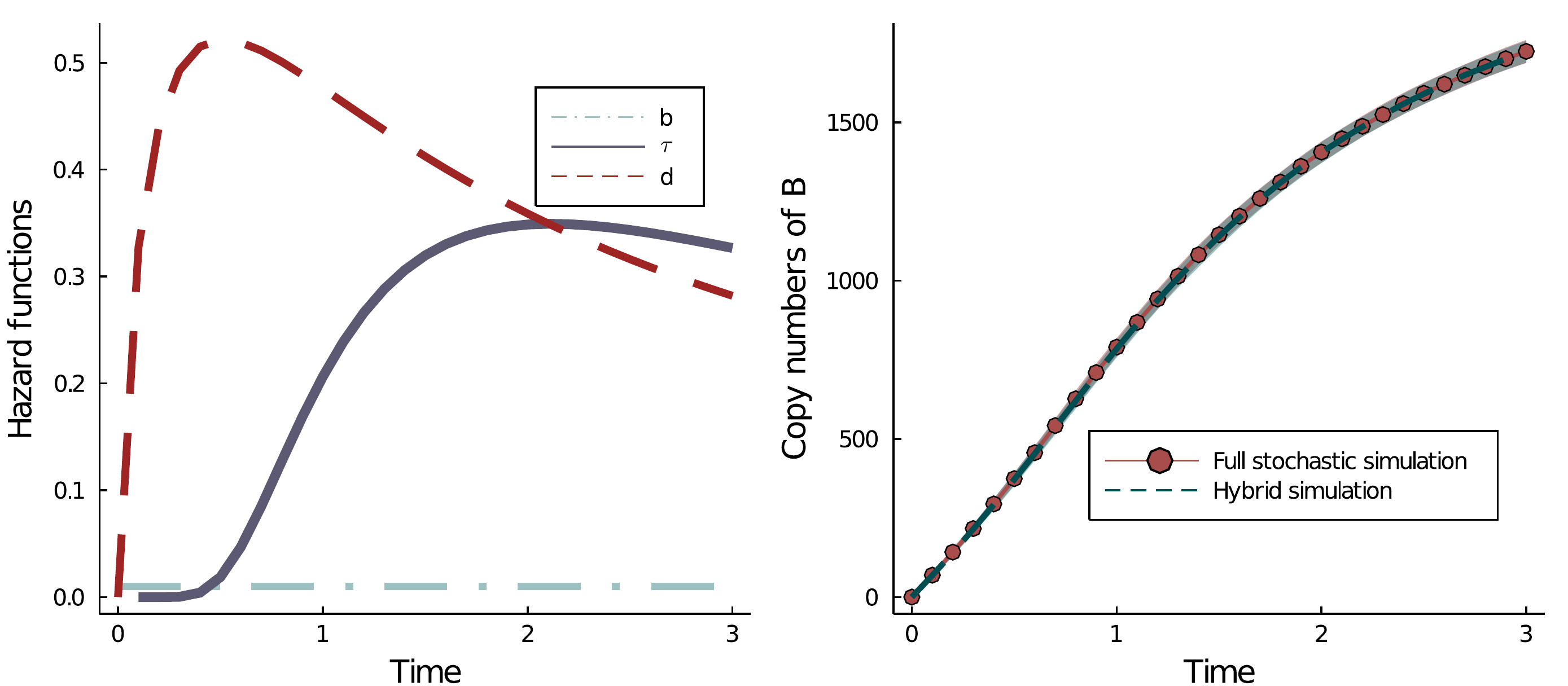}
    \caption{\label{fig:hybrid_example}%
    An example of the hybrid simulation approach. (Left) The shapes of the three hazard functions the \ac{CRN} described by \Cref{eq:model0}. Here, the birth rate 
    $b=0.01$. The distributions characterized by $\tau$, and $d$ are inverse gamma distribution with parameters 
    $(1.75, 4.25)$ and a Beta prime distribution with parameters $(1.75,1.25)$. The value of $n$ in this example is 
    $5000$. The full stochastic simulation took 305.714216 seconds, while the hybrid simulation took only 62.093832 seconds on a 2.3 GHz 18-Core Intel Xeon W machine.}
  \end{figure}

\subsection{Fast hybrid simulation}
Consider a situation when the species $A$ is abundantly available at the beginning of the reaction. Naturally, we 
expect the \ac{PDE} approximation to the age density of the species $A$ to be quite accurate, even though there will be 
considerable stochastic fluctuations in the copy numbers of $B$, at least initially. However, if we  approximate 
the age density of $A$ by the limiting \ac{PDE}, we can also approximate the initial growth of the $B$ molecules by 
a Poisson process whose time-varying intensity is driven by the \ac{PDE}. We use this idea to devise a hybrid 
simulation algorithm, which is, again, essentially an adaptation of the Doob--Gillespie's \ac{SSA} in the sense that 
it only draws next reaction times from an exponential distribution whose mean depends on the 
solution to the \ac{PDE}. For the sake of completeness, a pseudocode describing the idea is provided in \Cref{alg:hybrid}.

\begin{algorithm}[h]
  \small\topsep=0in\itemsep=0in\parsep=0in
    \begin{algorithmic}[1]
      \caption{\label[algorithm]{alg:hybrid}%
    \small   Pseudocode for the hybrid simulation algorithm}
    \Require $n$, $y_A$, $K$ (maximum number of reactions)
    \Ensure $t_1, t_2, \ldots$ (Timings of creation of $B$ molecules)
  \State Set $t_0=0$
  \For{$i=1,2,\ldots,K$}
  \State Calculate $\Lambda = \left(\int_0^\infty \tau(s) \, n y_{A}(t_{i-1}, s) \differential{s} \right)^{-1}$.
  \If{$0 < \Lambda <\infty$}
  \State Draw an exponential random variable $T$ with mean $\Lambda$, \ie, $T \sim \sys{Exponential}(\Lambda)$
  \State Set  $t_i = t_{i-1} +T$
  \Else \State Stop and break loop \EndIf
  \State Set $i=i+1$. 
  \EndFor
    \end{algorithmic}
  \end{algorithm}

\begin{figure}[ht]
  \centering 
  \includegraphics[width=0.75\linewidth]{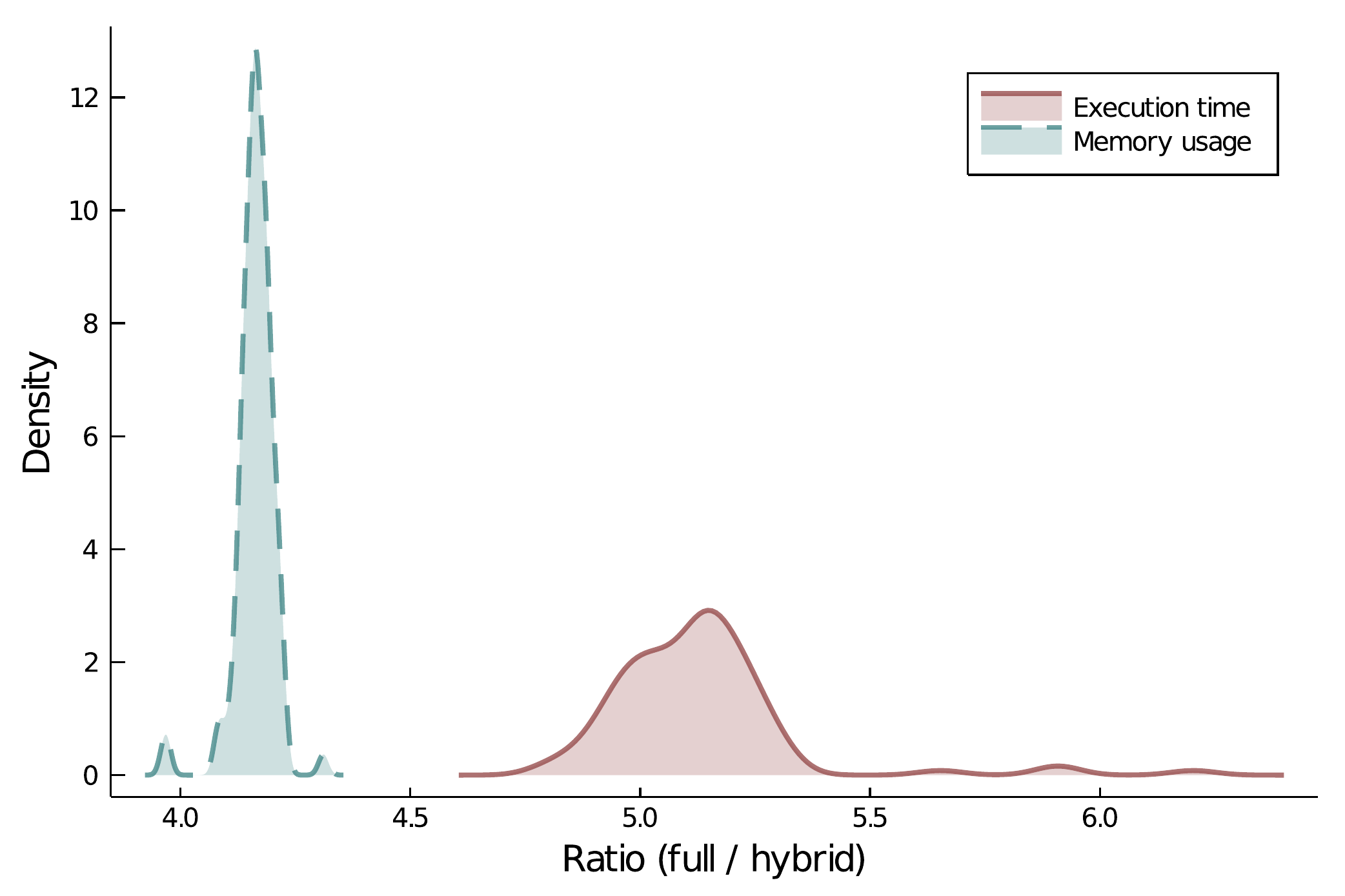}
  \caption{\label{fig:efficieny}%
  Efficiency of the hybrid simulation algorithm. The figure shows the empirical density of the ratios of 
  execution times, and 
  memory usage of the full stochastic simulation and those of the hybrid simulation algorithm described in 
  \Cref{alg:hybrid}. It is evident that the hybrid simulation algorithm is at least five times faster in terms of execution times, and at least 
  four times more efficient in terms of memory usage. The simulation set-up 
  is the same as \Cref{fig:hybrid_example}. The performance evaluation of the hybrid simulation is done using the 
  \emph{BenchmarkTools.jl} package \cite{BenchmarkTools.jl-2016} in Julia language \cite{Julia2017}
   }
\end{figure}

In \Cref{fig:hybrid_example}, we show the accuracy of the hybrid simulation algorithm. Expectedly, the hybrid 
simulation  is considerably  faster than the full stochastic simulation of the \ac{CRN} in \Cref{eq:model0}. A more elaborate  
comparison of performance is shown in \Cref{fig:efficieny}. However, it is worth noting that the hybrid simulation algorithm, by design, will underestimate the variance in the counting process 
for the species $B$. Therefore, one should use the hybrid simulation  when it suffices to get the mean trajectory 
accurately. Alternatively, one can borrow  ideas to estimate the variance correctly in other  simulation algorithms~\cite{Franz:2013:MRD,Harrison:2016:HAC,Kang:2019:MSR}. Similar 
ideas to expedite simulations have been proposed previously. For instance, the authors in \cite{arnab2015jump_diffusion} propose a jump-diffusion approximation to the stochastic \acp{CRN} and provide error analysis while others in \cite{Hepp:2015:AHS,Gupta:2019:SAM} introduce hybrid simulation methods using a piecewise deterministic Markov process. 

\section{Michaelis--Menten enzyme-kinetic reactions}
\label{sec:enzyme}
Michaelis--Menten enzyme-catalyzed chemical reactions form an important class of \acp{CRN} particularly
because of their vast applications in the industry \cite{CornishBowden2012Enzyme,Segel1975Enzyme}. Several descriptions of 
this class of reactions are available in the literature. For the sake of simplicity, in what 
follows we shall
adopt the simplest form of the Michaelis--Menten enzyme-catalyzed reactions. In this form, the \ac{CRN} comprises a reversible 
binding of a molecule of a substrate ($S$) and a molecule of an enzyme ($E$) into a molecule of a substrate-enzyme 
complex, and an irreversible conversion of a molecule of the complex into a molecule of a product ($P$) leaving the 
molecule of the enzyme free. That is, the system consists of the following reactions:
\begin{equation}
  \eqalign{E + S & \reactionrate{k_1}  C \eqcomma \\
  C & \reactionrate{k_{-1}}  E + S \eqcomma \\
  C & \reactionrate{k_2}  P + E \eqstop
  }
   \label{eq:enzyme_kinetic_reaction}
\end{equation}
In classical analysis, the quantities $k_1, k_{-1}$, and $k_2$ are reaction rate constants. When modelled 
stochastically using a \ac{CTMC}, the mean-field limit of the scaled concentrations is described by the following 
set of \acp{ODE} (see \cite{Kang2019QSSA} for more details):
\begin{equation}
  \eqalign{
    \timeDerivative{[E]} &= - k_1 [E] [S] + (k_{-1} + k_2) [C] \eqcomma \\
    \timeDerivative{[S]} &= - k_1 [E] [S] + k_{-1} [C] \eqcomma  \\
    \timeDerivative{[C]} &= k_1 [E] [S] - (k_{-1} + k_2) [C] \eqcomma  \\
    \timeDerivative{[P]} &= k_2 [C] \eqstop 
  }
  \label{eq:enzyme_mean_field_ode}
\end{equation}
The \ac{ODE} system in \Cref{eq:enzyme_mean_field_ode} has been studied extensively in the literature. We will 
adopt our measure-valued  representation to incorporate potential age structure in the Michaelis--Menten \ac{CRN}. 

\subsection{Enzyme kinetics with age structure}
We assume the  binding reaction depends on the age of the participating molecule of the enzyme. That is, only $k_1$ is 
age-dependent; $k_{-1}$, and $k_2$ are constants. For the species $E, S, C$, and $P$, define the measure-valued stochastic processes 
\begin{equation*}
  X_t^E \defeq \sum_{i=1}^{N_E(t)} \delta_{e_i(t)} \eqcomma 
  \quad X_t^S \defeq  \sum_{i=1}^{N_S(t)} \delta_{s_i(t)} \eqcomma  
  \quad X_t^C \defeq  \sum_{i=1}^{N_C(t)} \delta_{c_i(t)} \eqcomma  
  \quad X_t^P \defeq  \sum_{i=1}^{N_P(t)} \delta_{p_i(t)} \eqcomma  
\end{equation*}
where $N_E, N_S, N_C, N_P$ denote the counts of molecules of $E, S, C$, and $P$ respectively. Similarly, $e_i, s_i, c_i, p_i$ 
denote the age of the $i$-th molecule of $E, S, C$, and $P$ respectively. The process $X \defeq (X^E, X^S, X^C, X^P)$ is a 
Markov process on the space $D([0,T], \mathcal{M}_{P}(\setOfPositiveReals)^4)$. Please note that we need to scale 
the hazard function $k_1$ corresponding to the bimolecular reaction by $n^{-1}$ following the stochastic law of 
mass actions \cite{Ball:2006:AAM}.

As before, we are interested in 
the large-volume limit of the scaled process $n^{-1} X_t$. We expect $n^{-1} X_t$ to 
converge to a deterministic function $x_t \defeq (x_t^E, x_t^S, x_t^C, x_t^P)$ whose components 
$x_t^E, x_t^S, x_t^C, x_t^P$ are finite  measures on $\setOfPositiveReals$ by virtue of the \ac{LLN}.  Rigorous arguments 
supporting this convergence will be provided elsewhere.

Let $y_{E}$ denote the density  of the 
measure $x_t^E$ with respect to the Lebesgue measure. Also, let $y_S, y_C, y_P$ denote the concentrations of the 
$S, C$, and $P$ molecules. Then,  we get the following limiting system:
\begin{equation}
  \eqalign{
    \left(\partial_{t} + \partial_{s}\right) y_{E}(t,s) &={} - k_{1}(s) y_{E}(t,s) y_{S}(t) \eqcomma \\
    \timeDerivative{y_{S}(t)} &= {} - y_{S}(t) \int_0^\infty k_{1}(s) y_{E} (t,s) \differential{s}  
    + k_{-1} y_{C}(t) \eqcomma  \\
    \timeDerivative{y_{C}(t)} &={} y_{S}(t) \int_0^\infty k_{1}(s) y_{E} (t,s) \differential{s}  
    - (k_{-1} + k_2) y_{C}(t) \eqcomma  \\
    \timeDerivative{y_{P}(t)} &={} k_2 y_{C}(t) \eqcomma 
  }
  \label{eq:mm_pde_limit}
\end{equation}
with the boundary condition
\begin{equation*}
  y_{E}(t, 0) = (k_{-1} + k_2) y_{C}(t)
\end{equation*}
and the initial condition $y_E(0,s) = f_{E}(s)$ such that $\int_0^\infty f_{E}(s) \differential{s} = [E_0]$. Appropriate initial conditions for 
$S, C$, and $ P$ are also assumed. This limiting system can now be used 
to study the effects of delay in the binding reaction. One interesting 
approximation that has been widely applied in the context of Michaelis--Menten enzyme 
kinetic reactions is what is known as a \acl{QSSA} \cite{Eilertsen2020QSSA}. There are many 
forms of \acp{QSSA}, namely, \ac{sQSSA}, \ac{tQSSA}, and  \ac{rQSSA}. Detailed analysis of 
any of the \acp{QSSA} is beyond the scope of the present work. For the purpose of illustration, we informally 
describe  an analogue of the \ac{sQSSA} here.

\subsection{The \acl{sQSSA}}
The \acp{QSSA} are a multiscale approximation of the Michaelis--Menten
enzyme-kinetic reactions. The basic assumption behind the \acl{sQSSA}  is that 
 the substrate–enzyme complex $C$ reaches its steady-state much quicker than 
 the other species. In the deterministic set-up, the approximation  is achieved by setting 
 $\timeDerivative{y_{C}(t)} = 0$ in \Cref{eq:mm_pde_limit}, which allows one 
 to work with a smaller system of \acp{ODE}. Several conditions for the validity of the \ac{sQSSA} have been 
 proposed in the literature. See \cite{Kang2019QSSA} for a detailed discussion. 

Following the deterministic approach in our case, we set $\timeDerivative{y_{C}(t)} = 0$ in \Cref{eq:mm_pde_limit} to get a reduced 
\ac{PDE} system that is analogous to 
 the \ac{sQSSA}. To be more precise, $\timeDerivative{y_{C}(t)} = 0$ yields 
 \begin{equation*}
   y_{C}(t) = \frac{ y_{S}(t) \int_0^\infty k_1(s) y_{E}(t,s) \differential{s}  }{ k_{-1} + k_2 } \eqcomma 
 \end{equation*}
 which further yields an approximate system 
 \begin{equation}
   \eqalign{
     \timeDerivative{y_{S}(t)} &= -\frac{k_2}{k_{-1} + k_2} y_S(t) \int_0^\infty k_1(s) y_{E}(t,s) \differential{s} \eqcomma  \\
     \timeDerivative{y_{P}(t)} &= \frac{k_2}{k_{-1} + k_2} y_S(t) \int_0^\infty k_1(s) y_{E}(t,s) \differential{s} \eqstop 
   }
   \label{eq:sQSSA}
 \end{equation}
Notice that $y_E$ can be partially solved in terms of $y_S$ and $y_C$. Therefore, the reduced 
system of \acp{ODE} in \Cref{eq:sQSSA}
is indeed autonomous and therefore, serves as an \ac{sQSSA} of the 
\ac{CRN} in \Cref{eq:enzyme_kinetic_reaction}.

 In the stochastic set-up, the \acp{QSSA} are obtained by means of the probabilistic 
 multiscaling techniques developed in \cite{Kang:2013:STM, Kang:2012:MAH}. The stochastic 
 and the deterministic \acp{QSSA} mostly agree with each other with some notable differences. Please 
 see \cite{Kang2019QSSA} for examples of discrepancies as well as more details on the methods. Here, for paucity of space,  we do not consider 
 the stochastic \acp{QSSA} or  possible 
  discrepancies  between stochastic and deterministic methods in the present age-structured models.


\section{Prokaryotic auto-regulation}
\label{section:gene_network}

As another example, we consider a simple genetic network 
with feedback. We apply our approach using an 
age-dependent measure-valued process to build a 
model for a simple prokaryotic auto-regulation with
 a time delay. We modify an auto-regulation mechanism in the prokaryote gene network in \cite{Wilkinson:2012:SMS} (Section 1.5.7). We simplify the example by approximating transcription and translation as a one-step process with a time delay and replacing repression of gene by a protein dimer to repression by a single protein instead. For other related examples for gene transcription and translation, see Section 2.1.1 in \cite{Anderson:2015:SAB} and \cite{Rempala:2006:SMG,Kim:2017:RSB,Cappelletti:2019:LAS,Cao:2019:APE}.

Consider a genetic network with a gene ($G$), a protein ($P$),
and a gene-protein complex ($C$). Gene activates production of 
protein following a hazard function $b_P$, and protein 
degrades following a hazard function $d_P$. Protein can reversibly bind to gene to form a complex with binding hazard $b_C$ and 
unbinding hazard  $d_C$. Since gene-protein 
complex cannot participate in the production of protein, this is auto-regulation of the gene by its complex. Schematically, the reactions are as follows:
\begin{equation}
  \eqalign{
    G &\reactionrate{b_P} P+G \eqcomma \\
    P+G &\reactionrate{b_C} C \eqcomma \\
    C & \reactionrate{d_C} P+G \eqcomma \\
    P &\reactionrate{d_P} \emptyset \eqstop
  }
  \label{eq:gene}
\end{equation}
In (\ref{eq:gene}), we assume that the age of the gene is important. Therefore, the hazard functions 
$b_P$, and $b_C$ are assumed to be age-dependent whereas $d_C$ and $d_P$ are 
assumed to be constants. Note that after unbinding of the gene-protein complex, the age of gene is reset to zero. On the other hand, the age of gene is 
not affected by the protein production.

Denote by $N_G(t)$, $N_P(t)$, and $N_C(t)$  the total molecular counts of the gene, the protein, and the gene-protein complex at time $t$, respectively. For 
the species $G, P$, and $C$, define the measure-valued processes 
\begin{equation*}
  X_t^G \defeq \sum_{i=1}^{N_G(t)} \delta_{g_i(t)} \eqcomma 
  \quad X_t^P \defeq  \sum_{i=1}^{N_P(t)} \delta_{p_i(t)} \eqcomma  
  \quad X_t^C \defeq  \sum_{i=1}^{N_C(t)} \delta_{c_i(t)} \eqcomma  
\end{equation*}
where we denote the age of the $i$-th molecule of the species 
$G, P$, and $C$ by $g_i, p_i$, and $c_i$ respectively. As in the case of the Michaelis--Menten enzyme kinetic reaction, we 
scale the hazard function $b_{C}$ corresponding to the bimolecular reaction by $n^{-1}$ following the 
stochastic law of mass actions \cite{Ball:2006:AAM}.

The \ac{LLN} limit of the scaled process $n^{-1}X$ can be 
derived following by now familiar arguments of the previous examples. As one 
would expect, the scaled process $n^{-1}X$ converges to a 
deterministic function $x_t \defeq (x_t^G, x_t^P, x_t^C)$ whose 
components are finite measures on $\setOfPositiveReals$. Since we assume 
only the age of the gene is important, we consider the limiting age density  $y_G$ of the 
gene, which we obtain as the density, when it exists,  of the 
measure $x_t^G$ with respect to the Lebesgue measure. Similarly, define the limiting concentrations 
of the product $y_P$, and the complex $y_C$. The limiting system is then described by 
\begin{equation}
  \eqalign{
\left(\partial_{t} + \partial_{s}\right) y_{G}(t,s) &= - {b}_C(s)\,  y_{G}(t,s) y_{P}(t) \eqcomma \\
\timeDerivative{y_{P}(t)} & = \int_0^\infty b_{P}(s) y_{G}(t,s) \differential{s} - y_{P}(t) \int_0^\infty {b}_{C}(s) y_{G}(t,s) \differential{s} \\
&\quad +d_{C}\, y_{C}(t) -d_{P}\, y_{P}(t) \eqcomma \\
\timeDerivative{y_{C}(t)} & = y_{P}(t) \int_0^\infty {b}_{C}(s) y_{G}(t,s) \differential{s} -d_{C}\, y_{C}(t)  \eqcomma
}
\label{eq:gene_pde}
\end{equation}
with the boundary condition 
\begin{eqnarray*}
y_{G}(t,0) &=& d_{C}\, y_{C}(t)
\end{eqnarray*}
and the initial condition $y_{G}(0,s)=f_{G}(s)$, which specifies  the 
initial ages of the gene. Note that the hazard function for  unbinding of the gene-protein complex appears 
in the boundary condition since we assumed that the age of the  gene is 
reset to zero when the complex breaks into the gene and the protein. Also, recall 
 that $b_{P}(s)$ encodes a time delay in transcription and translation. For
  example, we may set $b_{P}(s)=r 1_{[\tau,\infty)}(s)$, which asserts that 
   protein is produced only when the age of gene is greater than $\tau$ with a hazard function $r$. 

\section{Discussion}
\label{sec:discussion}




Many biological processes with time delays, including \acp{CRN}, cannot be directly modeled using \acp{CTMC} due to non-exponentially distributed 
inter-event times of the processes. The simulation and analysis of  systems with an age structure and time delays become challenging since the system dynamics are affected by the inherent randomness (stochasticity) as well as time delays. One way to simulate such stochastic systems with age structure and 
time delays is to modify simulation algorithms for \ac{CTMC} models where the next reaction time and type are determined based on molecule counts of reactants. Bratsun et al.~\cite{Bratsun:2005:DSO}, Barrio et al.~\cite{Barrio:2006:ORH}, and Cai~\cite{Cai:2007:ESS} constructed 
modified \ac{SSA}s, while Anderson~\cite{Anderson:2007:MNR} introduced a modified next reaction method to simulate discrete stochastic chemical reaction networks with delays. Notably, all of those works assume that the time lags in the 
delayed reactions are constant.

\acp{CRN} with an age structure and randomly distributed time delays provide a more realistic  description of stochastic biophysical or chemical systems compared to the ones with fixed time delays. Unfortunately, the literature on stochastic systems with randomly distributed delays 
remains sparse. In a previous work by Koyama (Chapter 4 in~\cite{Koyama:2013:ASM}), the author investigated a stochastic kinetic network with a randomly distributed time delay 
where a delayed reaction can be interrupted by another reaction and can fail to complete. In another work by Marquez-Lago et al.~\cite{Marquez:2010:PDT}, the authors utilized randomly distributed time delays to incorporate spatial effects such as diffusion or translocation of molecules in temporal stochastic models. In a recent work by Choi et al.~\cite{Choi:2020:BID}, the authors described protein production in transcription and translation as a birth and death process with a randomly distributed time delay. 

In this paper, we developed a new way to incorporate an age structure and time delays in \acp{CRN} 
using age-dependent processes. We availed ourselves of  
 previous theoretical works \cite{Fournier2004microscopic,Tran2008limit,Tran2009traits,Meleard:2012:SFS} designed to study age-dependent population 
 dynamics. We applied those stochastic models in the context of \acp{CRN} to account for the  non-Markovian property due to the time delays. The use 
 of age-dependent hazard functions not 
 only enables us to model age-dependent time delays 
 or reaction rates but also covers the modeling of constant and random time delays in the
  existing literature. We illustrated our method using 
  simple biophysical systems in gene regulation and enzyme kinetics, but it will easily apply to 
  general \acp{CRN}. 

One potential disadvantage of the age-dependent processes is that simulation can be prohibitive since 
the age of each individual molecule of the chemical species of 
interest needs to be tracked over the entire simulation time. Therefore, we 
derived a large-volume limit of the age-dependent process for \acp{CRN} in the form of \acp{PDE} 
using the analytic methods in \cite{Fournier2004microscopic,Tran2008limit,Tran2009traits,Meleard:2012:SFS} and used the \ac{PDE} limit to construct 
a hybrid simulation algorithm, which, in our example, turned 
out to be five times faster than the full stochastic simulation. Moreover, we approximated a \acl{MFPT} efficiently utilizing the theoretical limit.


In this work, we emphasized how age-structured processes and their large-volume limits can be applied
 to model \acp{CRN}, in particular, biophysical or chemical systems with time delays. Many 
 previous findings for general \acp{CRN} under Markovian 
 assumption can be reinvestigated and extended to non-Markovian settings using age-structured 
 processes. It would be interesting to see how the long time behavior of stochastic \acp{CRN} is affected by 
 incorporating age structure.   For example, it would be 
 interesting to study stationary distributions of autocatalytic \acp{CRN} with 
 switching behavior~\cite{Bibbona:2020:SDS}, to identify a class of \acp{CRN} maintaining product-form Poisson 
 distributions for all times~\cite{Anderson:2020:TPP}, and to find when \acp{CRN} show 
 nonexplosive behavior~\cite{Anderson:2018:NSM}. Another interesting direction will be  to study stability of \acp{CRN}~\cite{Agazzi:2018:SSC} and to estimate transition times between different attractors in \acp{CRN}~\cite{Agazzi:2018:LDT}. 

 For the sake of simplicity, we have assumed in this 
 paper that the molecular entities of all chemical species are abundant 
 at the same order of magnitude so as to obtain the
 large-volume limit under the classical scaling. A natural extension 
 of this work is to consider general \acp{CRN} with a wide range of molecular abundances 
 and reaction rates where we can apply multiscale approximations to reduce model complexity~\cite{Ball:2006:AAM,Kang:2013:STM,Kang:2014:CLT}. We leave 
 such investigation to future work. In this paper, we briefly described how an analogue of \ac{QSSA} can be derived in the Michaelis-Menten enzyme-kinetic reactions. As shown in the related previous work~\cite{Kang2019QSSA,Eilertsen2020QSSA}, both deterministic and stochastic \acp{QSSA} can be revisited with an extension of our approach to multiscale approximations in enzyme kinetics under non-Markovian setting. 
Another promising application of our approach seems to be in 
parameter inference and survival analysis of general \acp{CRN} with age structure. Given the current interests in 
pandemic modeling, such \acp{CRN} could lead to interesting examples in population dynamics and epidemiology. 
We hope to be able to pursue such work in the near future.


\begin{appendices}
\renewcommand{\theequation}{\thesection.\arabic{equation}}

\crefalias{section}{appsec}

\section{Table of symbols}
\label[Appendix]{appendix:tableofsymbols}

\begin{table}[H]
  \centering
  \begin{tabular}{|| c | c ||}
    \hline 
    Symbol & Meaning \\
    \hline 
    $\setOfNaturals$ & The set of natural numbers\\
    $\setOfReals$ & The set of reals \\
    $\setOfPositiveReals$  & The set of non-negative reals\\
    $\indicator{A}(x)$ & Indicator (characteristic) function of the set $A$\\
    $\delta_x$ & Dirac delta function at $x$\\
    $\borel{A}$ & The Borel $\sigma$-field of subsets of a set $A$\\
    $\mathcal{M}_{P}(E)$ & The space of finite point measures on the set $E$\\
    $D([0,T], E)$ & The space of  $E$-valued \cadlag functions defined on $[0,T]$\\
    $\measureIntegral{\mu}{f}$ & The integral $\int f \differential{\mu}$ \\
    \hline 
  \end{tabular}
\end{table}

\section{Acronyms}\label{appendix:acronym}

\begin{acronym}[OWL-QN]
	\acro{ABM}{Agent-based Model}
	\acro{ADMM}{Alternating Direction Method of Multipliers}
	\acro{BA}{Barab\'asi-Albert}
	\acro{BCS}{Bioinspired Communication Systems}
	\acro{BD}{Birth-death}
	\acro{BM}{Brownian Motion}
	\acro{CBQA}{Cost-Based Queue-Aware}
	\acro{CBS}{Cost-Based Scheduling}
	\acro{CCDF}{Complementary Cumulative Distribution Function}
	\acro{CDC}{Centers for Disease Control and Prevention}
	\acro{CDF}{Cumulative Distribution Function}
	\acro{CDN}{Content Distribution Network}
	\acro{CIM}{Conditional Intensity Matrix}
	\acro{CLT}{Central Limit Theorem}
	\acro{CM}{Configuration Model}
	\acro{CME}{Chemical Master Equation}
	\acro{CoM}{Compartmental Model}
	\acro{CRC}{Collaborative Research Centre}
	\acro{CRM}{Conditional Random Measure}
	\acro{CRN}{Chemical Reaction Network}
	\acro{CTBN}{Continuous Time Bayesian Network}
	\acro{CTMC}{Continuous Time Markov Chain}
	\acro{DCFTP}{Dominated Coupling From The Past }
	\acro{DFG}{German Research Foundation}
	\acro{DTMC}{Discrete Time Markov Chain}
  \acro{DRC}{Democratic Republic of Congo}
	\acro{DSA}{Dynamic Survival Analysis}
	\acro{ECMP}{Equal-cost Multi-path routing}
	\acro{EDF}{Earliest Deadline First}
	\acro{ER}{Erd\"{o}s-R\'{e}nyi}
	\acro{ESI}{Enzyme-Substrate-Inhibitor}
	\acro{FCFS}{First Come First Served}
	\acro{FCLT}{Functional Central Limit Theorem}
	\acro{FIFO}{First In First Out}
	\acro{FJ}{Fork-Join}
	\acro{FPT}{First Passage Time}
	\acro{GBP}{General Branching Process}
	\acro{HJB}{Hamilton–Jacobi–Bellman}
	\acro{ID}{Information-Dissemination}
	\acro{iid}{independent and identically distributed}
	\acro{IoT}{Internet of Things}
	\acro{IPS}{Interacting Particle System}
	\acro{IT}{Information Technology}
	\acro{JIQ}{Join-Idle-Queue}
	\acro{JMC}{Join the Minimum Cost}
	\acro{JSQ}{Join the Shortest Queue}
	\acro{KL}{Kullback-Leibler}
	\acro{LDF}{Latest Deadline First}
	\acro{LDP}{Large Deviations Principle}
	\acro{LLN}{Law of Large Numbers}
	\acrodefplural{LLN}[LLNs]{Laws of Large Numbers}
	\acro{LNA}{Linear Noise Approximation}
	\acro{MABM}{Markovian Agent-based Model}
	\acro{MAKI}{Multi-Mechanism Adaptation for the Future Internet}
	\acro{MAPK}{Mitogen-activated Protein Kinase}
	\acro{MCMC}{Markov Chain Monte Carlo}
	\acro{MDS}{Maximum Distance Separable}
	\acro{MFPT}{Mean First Passage Time}
	\acro{MGF}{Moment Generating Function}
	\acro{MLE}{Maximum Likelihood Estimate}
	\acro{MM}{Michaelis-Menten}
	\acro{MPI}{Message Passing Interface}
	\acro{MPTCP}[Multi-path TCP]{Multi-path Transmission Control Protocol}
	\acro{MTM}{Mass Transfer Model}
	\acro{MSE}{Mean Squared Error}
	\acro{ODE}{Ordinary Differential Equation}
	\acro{P2P}{Peer-to-Peer}
	\acro{PDE}{Partial Differential Equation}
	\acro{PDF}{Probability Density Function}
	\acro{PGF}{Probability Generating Function}
	\acro{PGM}{Probabilistic Graphical Model}
	\acro{PMF}{Probability Mass Function}
	\acro{PPM}{Poisson Point Measure}
	\acro{PRM}{Poisson Random Measure}
	\acro{psd}{positive semi-definite}
	\acro{PT}{Poisson-type}
	\acro{QoE}{Quality of Experience}
	\acro{QoS}{Quality of Service}
	\acro{QSSA}{Quasi-Steady State Approximation}
\acro{RBM}{Reflecting Brownian Motion}
	\acro{rQSSA}{reversible QSSA}
	\acro{SAN}{Stochastic Automata Network}
	\acro{SD}{Standard Deviation}
	\acro{SDS}{Survival Dynamical System}
	\acro{SEIR}{Susceptible-Exposed-Infected-Recovered}
	\acro{SI}{Susceptible-Infected}
	\acro{SIR}{Susceptible-Infected-Recovered}
	\acro{SIS}{Susceptible-Infected-Susceptible}
	\acro{SPDE}{Stochastic Partial Differential Equation}
	\acro{sQSSA}{standard QSSA}
	\acro{SRBM}{Semi-martingale Reflecting Brownian Motion}
	\acro{SRPT}{Shortest Remaining Processing Time}
	\acro{SSA}{Stochastic Simulation Algorithm}
	\acro{ssLNA}{Slow-scale Linear Noise Approximation}
	\acro{STC}{Stone Throwing Construction}
	\acro{TCP}{Transmission Control Protocol}
	\acro{tQSSA}{total QSSA}
	\acro{WS}{Watts-Strogatz}
	\acro{whp}{with high probability}
	\acro{WSU}{Washington State University}
	\acro{RAM}{Robust Adaptive Metropolis}
	\acro{ASM}{Adaptive Scaling Metropolis}
\end{acronym}

%
%

\section{Software}
The numerical results in this paper are obtained by the Julia programming language \cite{Julia2017}. The Julia scripts (compatible with version 1.4.1) used 
in this paper have been made available publicly at a dedicated GitHub repository \cite{delaymodel2020}. 




\section*{Funding}
WKB was supported by the National Institute of Allergy and Infectious Diseases (NIAID) 
grant R01 AI116770, the National Science Foundation (NSF) grant DMS-2027001 and the Ohio State University's President's Postdoctoral Scholars Program (PPSP). EK was supported by NIAID grant 
R01 AI116770 and the NSF grants DMS-2027001 and DMS-1853587. GAR was supported by the NSF grants DMS-2027001 and DMS-1853587. HWK was supported in 
part by the NSF under the grant DMS-1620403. The
 project was initiated when HWK was visiting the Mathematical Biosciences Institute (MBI) 
 at the Ohio State University in Summer 2019. The authors acknowledge the hospitality and the support of MBI. The
content of this manuscript is solely the responsibility of the authors and does not represent the
 official views of NSF, NIGMS, NIAID, or NIH.
\end{appendices}

\section*{References}

\bibliographystyle{iopart-num}
\bibliography{bibliography_WKB}

\end{document}